\documentclass[iop,revtex4]{emulateapj}

\usepackage{natbib}
\usepackage{pdflscape}

\usepackage{soul}
\usepackage[usenames,dvipsnames]{color}
\sethlcolor{YellowGreen}

\begin{document}

\title{The 2010 Eruption of the Recurrent Nova U Scorpii: The Multi-Wavelength Light Curve}
\author{Ashley Pagnotta\altaffilmark{1}, Bradley E. Schaefer\altaffilmark{2}, James L. Clem\altaffilmark{2}, Arlo U. Landolt\altaffilmark{2}, Gerald Handler\altaffilmark{3}, Kim L. Page\altaffilmark{4}, Julian P. Osborne\altaffilmark{4}, Eric M. Schlegel\altaffilmark{5}, Douglas I. Hoffman\altaffilmark{6}, Seiichiro Kiyota\altaffilmark{7}, Hiroyuki Maehara\altaffilmark{8}}
\altaffiltext{1}{Department of Astrophysics, American Museum of Natural History, New York, NY 10024, USA}
\altaffiltext{2}{Department of Physics and Astronomy, Louisiana State University, Baton Rouge, LA 70803, USA}
\altaffiltext{3}{Nicolaus Copernicus Astronomical Center, Bartycka 18, 00-716 Warsaw, Poland}
\altaffiltext{4}{Department of Physics and Astronomy, University of Leicester, Leicester LE1 7RH, UK}
\altaffiltext{5}{Department of Physics and Astronomy, University of Texas at San Antonio, San Antonio, TX 78249, USA}
\altaffiltext{6}{Infrared Processing and Analysis Center, California Institute of Technology, Pasadena, CA 91125, USA}
\altaffiltext{7}{7-1 Kitatasutomi, Kamagaya, Chiba 273-0126, Japan}
\altaffiltext{8}{Kwasan Observatory, Kyoto University, Yamashina-ku, Kyoto 607-8471, Japan}
\email{pagnotta@amnh.org}

\begin{abstract}
The recurrent nova U Scorpii most recently erupted in 2010. Our collaboration observed the eruption in bands ranging from the {\it Swift} XRT and UVOT {\it w2} (193 nm) to {\it K}-band (2200 nm), with a few serendipitous observations stretching down to {\it WISE} {\it W2} (4600 nm). Considering the time and wavelength coverage, this is the most comprehensively observed nova eruption to date. We present here the resulting multi-wavelength light curve covering the two months of the eruption as well as a few months into quiescence. For the first time, a U Sco eruption has been followed all the way back to quiescence, leading to the discovery of new features in the light curve, including a second, as-yet-unexplained, plateau in the optical and near-infrared. Using this light curve we show that U Sco nearly fits the broken power law decline predicted by Hachisu \& Kato, with decline indices of $-1.71 \pm 0.02$ and $-3.36 \pm 0.14$. With our unprecedented multi-wavelength coverage, we construct daily spectral energy distributions and then calculate the total radiated energy of the eruption, $E_\mathrm{rad}=6.99^{+0.83}_{-0.57} \times 10^{44}$ erg. From that, we estimate the total amount of mass ejected by the eruption to be $m_\mathrm{ej}=2.10^{+0.24}_{-0.17} \times 10^{-6} M_\odot$. We compare this to the total amount of mass accreted by U Sco before the eruption, to determine whether the white dwarf undergoes a net mass loss or gain, but find that the values for the amount of mass accreted are not precise enough to make a useful comparison.  
\end{abstract}


\section{U Scorpii}
\label{sec:intro}
The 2010 eruption of the recurrent nova (RN) U Scorpii was predicted by \citet{schaefer2005a} and discovered independently by B. G. Harris and S. Dvorak as part of an intense monitoring campaign coordinated by our group at Louisiana State University (LSU) and the American Association of Variable Star Observers (AAVSO) \citep{schaefer2010c, schaefer2010d, simonson2010a}. The discovery triggered a worldwide invocation of both pre-planned (target of opportunity) and serendipitous observing programs. This was the tenth observed eruption of U Sco \citep{schaefer2010b}, and became by far the best observed nova eruption to date. The eruption began on JD $2455224.32 \pm 0.12$, peaked on JD $2455224.69 \pm 0.07$ ($T_\mathrm{0}$) at $V=7.5$ mag, and returned to quiescence 67 days later \citep{schaefer2010d}.

We present multi-wavelength photometry of the complete eruption obtained by our extensive collaboration of both professional and amateur astronomers. U Sco was observed in all wavelengths from radio to gamma rays during the 2010 eruption, with detections from IR to soft X-ray. The fast time variations of the light curve are discussed in detail in \citet{schaefer2011b}; here we focus on the overall shape and spectral energy distribution (SED). We construct daily SEDs and use them to get a new estimate of the total amount of mass ejected during the eruption, $m_\mathrm{ej}$, following the method described in \citet{shara2010a}. It is critical to get a good measurement of $m_\mathrm{ej}$ and compare it to estimates of the total amount of mass accreted during the time interval preceding the eruption to ascertain whether the white dwarf (WD) in the system is gaining or losing mass over the course of the eruption cycle. If the WD, which is already near the Chandrasekhar limit \citep{hachisu2000a,thoroughgood2001a}, is in fact gaining mass and is composed primarily of carbon and oxygen, it must eventually become a Type Ia supernova (SN Ia). Determining whether RNe can become SNe Ia takes us one step closer to solving the long-standing SN Ia progenitor problem and improving the cosmological measurements that rely  on SN Ia standard candle distances.

\begin{figure*}
\centering
\epsscale{1.0}
\plotone{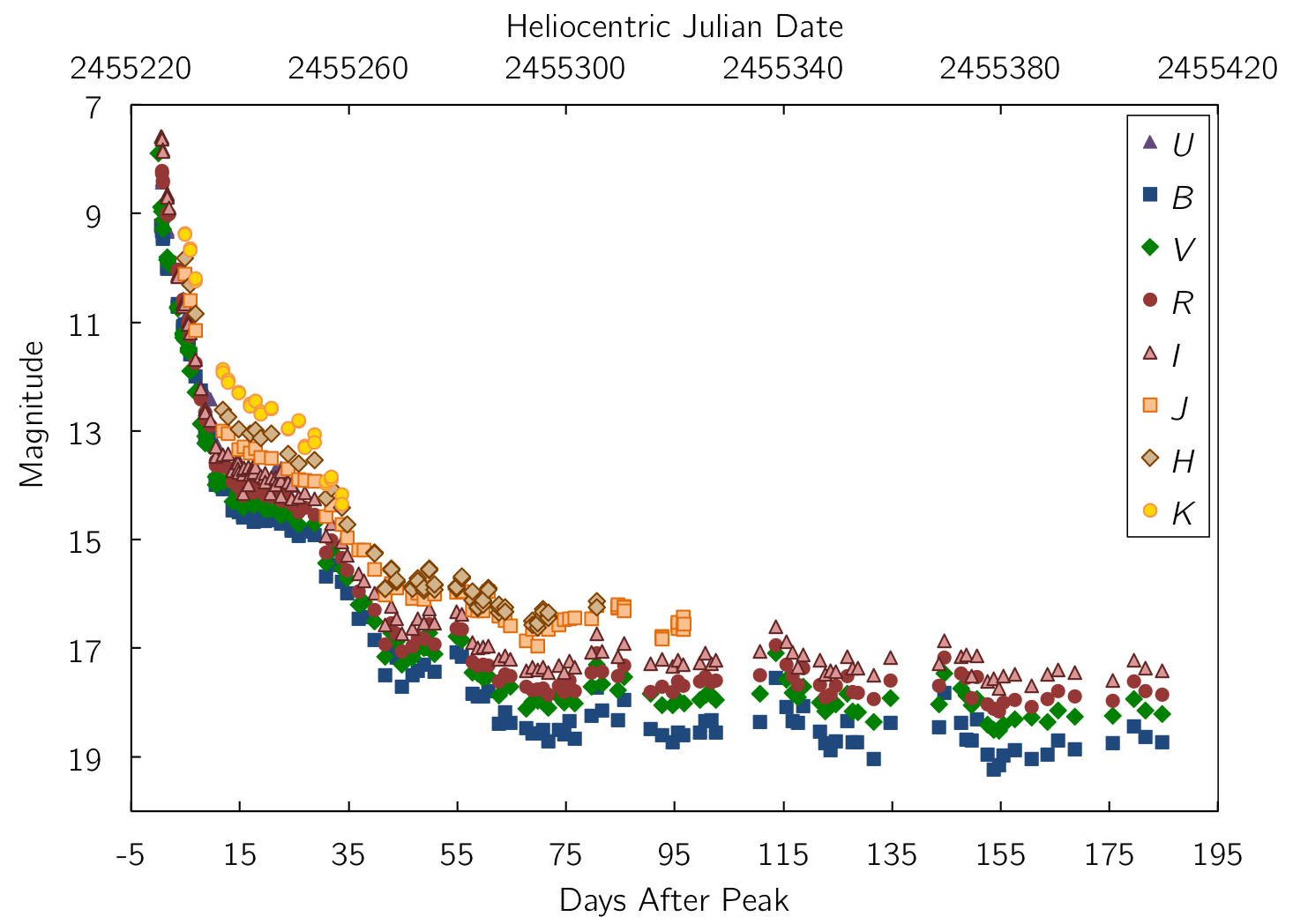}
\caption{U Sco {\it UBVR$_c$I$_c$JHK} Full Light Curve from the CTIO 1.3m and SAAO 0.5m Telescopes. A $U$-band filter was available for the SAAO 0.5m but not the CTIO 1.3m, so the $U$-band data cover only the first month. Due to decreasing signal-to-noise in the near-IR as U Sco faded, the {\it J}, {\it H}, and {\it K} observations end on days 97, 81, and 34 after peak, respectively. This figure shows all of our non-Str\"{o}mgren out-of-eclipse observations from CTIO and SAAO, lasting until 2010 July 31. The observations within a tenth of a phase of eclipse have been removed to better show the overall trend of the light curve. We followed U Sco far past its return to quiescence to ensure that all late-time phenomena would be observed.}
\label{fig:smartssaao}
\end{figure*}

\section{Observations}
\label{sec:observations}
A full list of observers and observation details for the multi-wavelength light curve in this paper is presented in Table \ref{tab:observers}. The observations were largely coordinated by our group at LSU and were carried out at telescopes all over the world and from a number of orbiting observatories. The discovery occurred on 2010 January 28 (with the first image taken by B. J. Harris at HJD 2455224.94) and, upon confirmation, we immediately notified our collaboration and began observing U Sco nearly constantly. The {\it BVR$_c$I$_c$JHK} observations by Pagnotta using the SMARTS 1.3m at CTIO and {\it UBVR$_c$I$_c$by} observations by Handler using the SAAO 0.5m form the backbone of our optical/near-IR light curve and can be seen in Figure \ref{fig:smartssaao}, where the observations within a tenth of a phase of eclipse have been removed to better show the overall trend. Regular observations were made by the {\it Swift} XRT and UVOT instruments and analyzed by Page on behalf of the {\it Swift} Nova-CV Group; they provide the UV and X-ray light curves presented here. U Sco was observed in the hard X-ray and gamma ray regime by INTEGRAL, the {\it Swift} BAT, RXTE, the Fermi GBM, and the MAXI detectors on the International Space Station, but was not detected in hard X-rays by any instrument (\citealp{manousakis2010a}, H. Krimm 2010, private communication, E. Kuulkers \& K. Mukai 2010, private communication, J. Rodi 2010, private communication, T. Mihara and the MAXI team 2015, private communication). The GMRT and ATCA observatories looked at U Sco in radio, but there were no positive detections (G. C. Anupama \& N. G. Kantharia 2010, private communication, and S. Eyres \& T. O'Brien 2010, private communication). Additionally, there were a few serendipitous observations made by the {\it WISE} satellite during its infrared survey mission.

\begin{figure*}
\centering
\epsscale{1.0}
\plotone{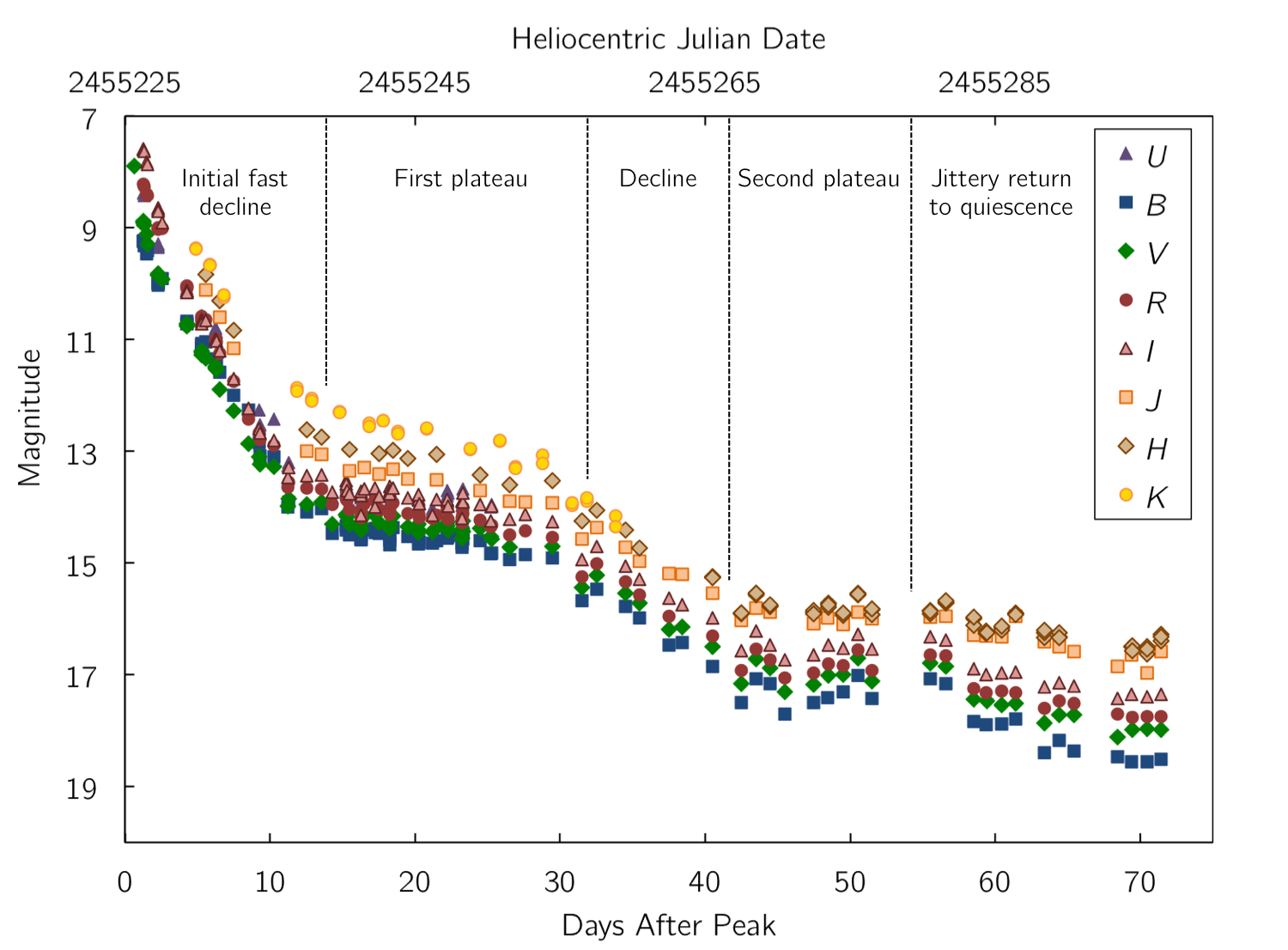}
\caption{Zoomed-in version of the U Sco {\it UBVR$_c$I$_c$JHK} Eruption Light Curve from the CTIO 1.3m and SAAO 0.5m Telescopes for phases 0.1-0.9. This plot ends at day 71 (after peak), showing just the actual eruption of U Sco and a few days of quiescence. The eruption light curve can be divided into five main parts: the initial fast decline (days 0-14), the first plateau (days 14-32), the subsequent decline (days 32-41), the second plateau (days 41-54), and the jittery return to quiescence (days 54-67). This marks the first time that U Sco has been systematically followed past day 30, and therefore the first time the second plateau has been observed \citep{pagnotta2010a}.}
\label{fig:eruption}
\end{figure*}

The U Sco 2010 optical/near-IR eruption light curve can loosely be broken down into five different parts: the initial fast decline (days $0{-}14$ after peak), the first plateau (days $14{-}32$), the subsequent fall (days $32{-}41$), the second plateau (days $41{-}54$), and the jittery return to quiescence (days $54{-}67$). Figure \ref{fig:eruption} shows a subset of the data from Figure \ref{fig:smartssaao}, namely the out-of-eclipse observations taken during the course of the actual eruption, to make it easier to distinguish the overall shape of the light curve as well as the different parts. The initial fast decline lasted for approximately two weeks, from the peak until the start of the plateau on day 14. The start of the plateau occurs at approximately the same time it did in previous eruptions and coincides with the detection of the supersoft X-rays by {\it Swift }and the reappearance of the optical eclipses \citep{schlegel2010b,schaefer2011b}. These three events are correlated with the thinning of the expanding nova shell, as described by \citet{hachisu2008a}: once the shell has become optically thin, it no longer blocks the actual U Sco binary (the two stars and the re-forming accretion disk), and both the X-rays and the eclipses are revealed. Some of the X-rays are reprocessed by circumstellar material to longer wavelengths, and this provides the source of the extra light that causes the UV, optical, and near-IR plateaus. The first plateau lasts for approximately 18 days, during which the eclipses are shallower than during quiescence and the secondary eclipse is visible \citep{schaefer2011b}. At the end of the plateau, the UV, optical, and near-IR light curves resume their drop-off, at a rate slower than the initial decline. 

The second plateau begins on approximately day 41 after peak and lasts until day 54. During this second plateau, the average $V$-band magnitude is 17.01, and it varies between 16.71 and 17.30, hovering approximately 1 mag above the quiescent level of $V=18.0$. This second plateau of U Sco was not seen in any previous eruption, because none of them were followed regularly past day 30. Concurrent fast time series photometry of U Sco showed aperiodic optical dips with amplitudes of up to 0.6 mag \citep{schaefer2011b}. These dips are linked to eclipses of the central light source due to turbulence in the still re-forming accretion disk, but we are unaware of any theoretical link between them and the second plateau. On approximately day 54 the second plateau ended and U Sco resumed fading, returning to quiescence on day 67.

The optical/near-IR colors are relatively constant throughout the eruption. The notable exception to this color constancy is shortly after peak: initially U Sco was brighter in $V$ than in $B$, but around HJD 2455228 this switched, and $B$ was brighter for approximately 7 days, until HJD 2455235 when the colors switched back and $V$ remained brighter than $B$ for the rest of the eruption, which is the usual state of the system in quiescence.

\begin{figure*}
\centering
\epsscale{1.0}
\plotone{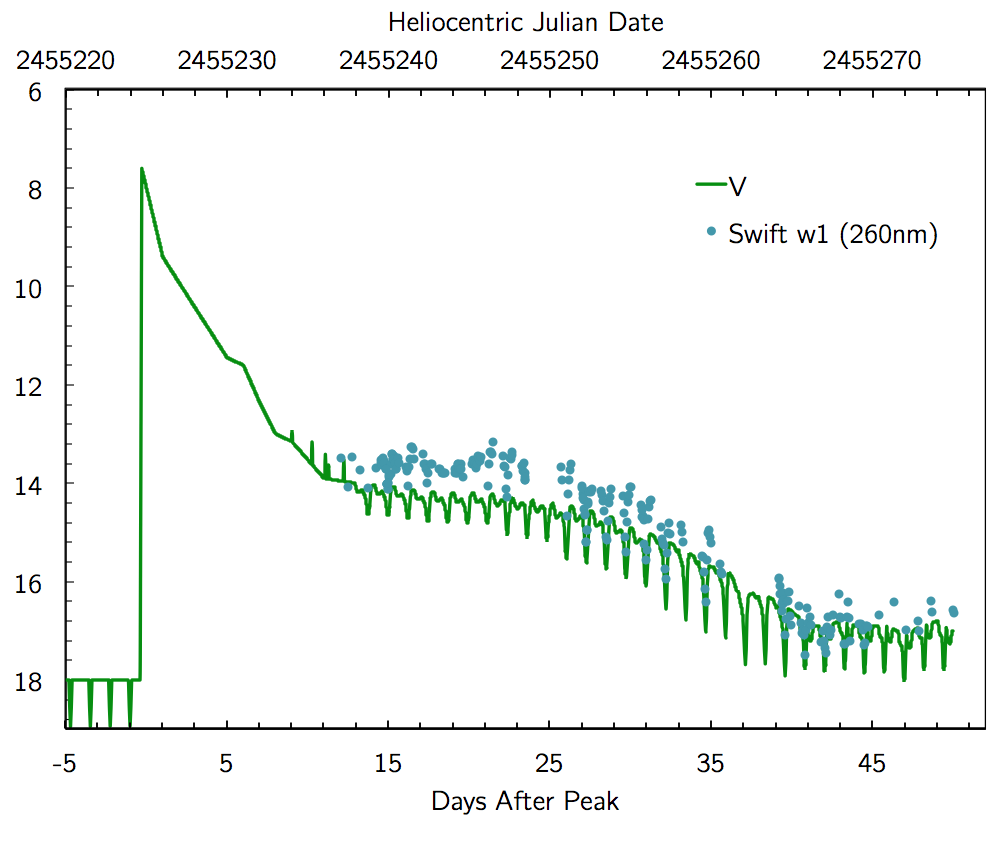}
\caption{UV vs. $V$-band comparison. The {\it Swift} $w1$ (260 nm) light curve is plotted along with the idealized $V$-band light curve \citep{pagnotta2010a}. The shape of the UV light curve closely tracks that of the optical, with the UV also showing primary eclipses on U Sco's orbital period.}
\label{fig:w1vband}
\end{figure*}

\begin{figure*}
\centering
\epsscale{1.0}
\plotone{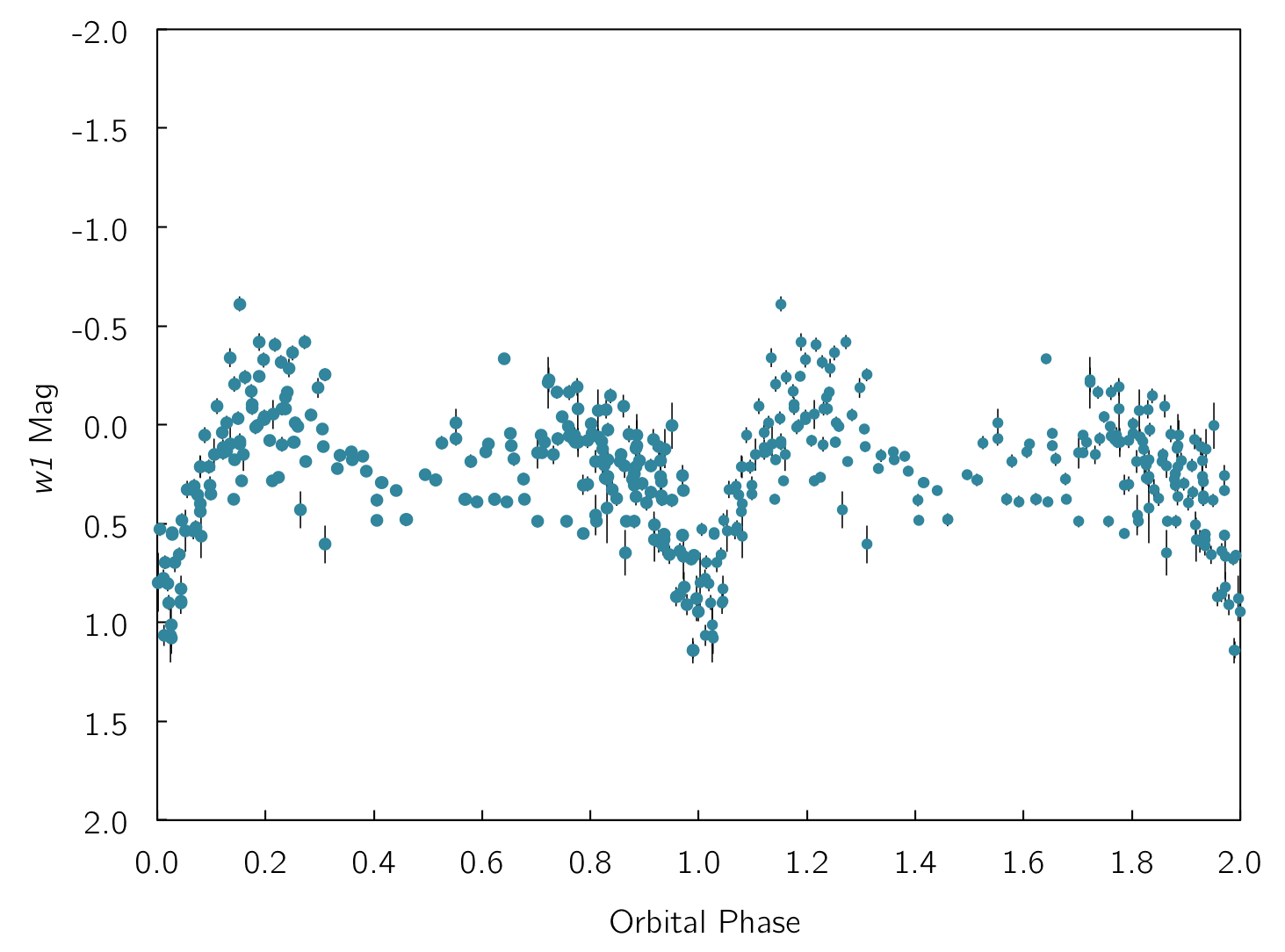}
\caption{Phased UV light curve. The {\it Swift} $w1$ light curve is folded on the known orbital period of U Sco ($P_{\mathrm orb} = 1.23$ days, \citealp{schaefer2010b}) and the phased light curve is plotted twice to make it easier to visually identify the primary eclipse, which occurs at phase 0. The error bars represent the 1$\sigma$ error; when they are not visible it is because the error is smaller than the point size. The points in the figure come from all times after the reappearance of the eclipses on Day 14 except during the second plateau (Days 41-54), during which the aperiodic optical dips (see Section \ref{sec:observations} and \citealp{schaefer2011b}) add noise to the phased light curve. The primary eclipse is clearly visible, and there appears to be indication of the secondary eclipse as well, although there is too much scatter to say definitively that it is detected.}
\label{fig:uvphased}
\end{figure*}

\begin{figure*}
\centering
\epsscale{0.8}
\plotone{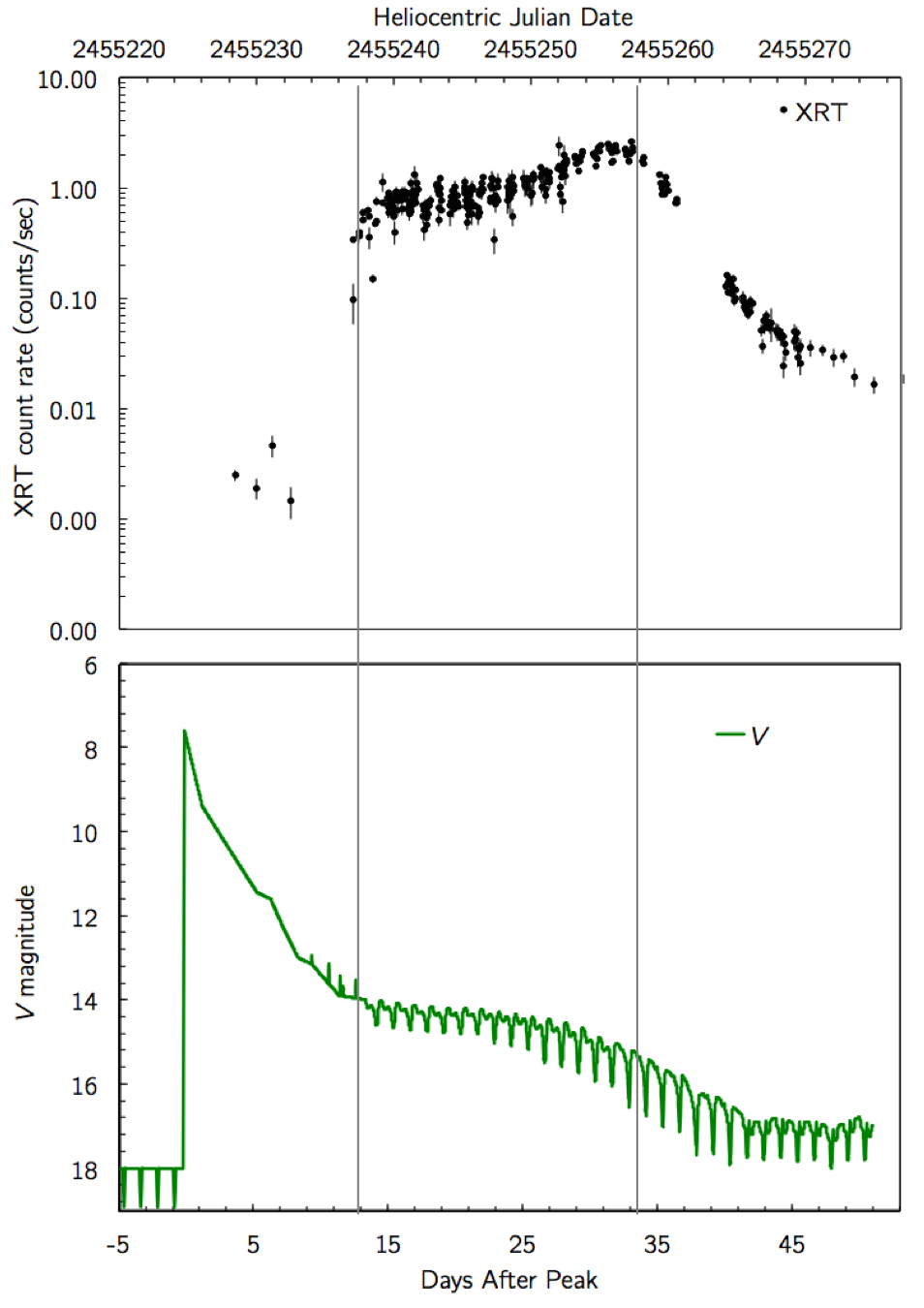}
\caption{{\it Swift} XRT+$V$-Band Light Curve. The {\it Swift} XRT light curve is shown in the top figure, and compared to the idealized $V$-band light curve, shown in the bottom \citep{pagnotta2010a}. The X-rays brighten drastically around day 14. At the same time, the optical plateau begins and the eclipses reappear. These three effects are all observed at the same time, when the expanding nova shell becomes optically thin and we can once again see down to the underlying binary system. The X-ray turn-on and peak are marked on this figure with solid grey vertical lines.}
\label{fig:xrtvband}
\end{figure*}

The {\it Swift} UVOT \citep{roming2005a} regularly monitored the eruption in the ultraviolet bands $w1$ (260.0 nm) and $w2$ (192.8 nm), and also took occasional snapshots in the $u$ (346.5 nm) and $m2$ (224.6 nm) bands. The shape of the UV light curve is very similar to that of the optical/near-IR light curve; this can be seen in Figure \ref{fig:w1vband}, which compares the {\it {\it Swift}} $w1$ light curve to the idealized $V$-band light curve, which was constructed from more than 35,000 fast time series observations \citep{pagnotta2010a}. The eclipses that are visible in the optical and near-IR data are visible in the UV as well, as was originally noted in \citet{ness2012a}. Figure \ref{fig:uvphased} shows a phased {\it Swift w1} light curve that clearly shows the primary eclipse and shows indications of the secondary eclipse, although there is a lot of scatter. The {\it Swift} XRT, which is sensitive to the energy range $0.3{-}10$ keV, also observed U Sco regularly. The X-ray light curve compared to the idealized $V$-band light curve can be seen in Figure \ref{fig:xrtvband}, as well as in \citet{ness2012a} and \citet{orio2013a}, which discuss different aspects of the X-ray emission. The eclipses start to reappear in the XRT light curve at approximately the same time they do in optical, near-IR, and UV, however they are not as well defined until around day 30 (\citealp{osborne2010a,schaefer2010f}; E. M. Schlegel et al. 2015, in preparation), perhaps due to the presence of aperiodic X-ray dips early on in the eruption \citep{ness2012a}.

The {\it WISE} infrared survey mission \citep{wright2010a} was operating during the eruption and serendipitously observed U Sco ten times from HJD 2455252.5 to 2455253.3 in bands $W1$ (3400 nm) and $W2$ (4600 nm). (For clarity, we use lowercase $w1$ and $w2$ for the {\it Swift} bands and uppercase $W1$ and $W2$ for the {\it WISE} bands throughout this paper.) We extracted these observations from the {\it WISE} All-Sky Single Exposure (L1b) Source Table\footnote{http://irsa.ipac.caltech.edu/cgi-bin/Gator/nph-dd?catalog=wise\_allsky\_4band\_p1bs\_psd}.

Table \ref{tab:mags} shows the average daily magnitudes in all bands, with missing days or observations that occurred during eclipse filled in using the interpolation and extrapolation methods described in Section \ref{sec:sed} and denoted in the tables with daggers ($\dagger$) and double daggers ($\ddagger$)  on the interpolated and extrapolated values, respectively. There are only two days with {\it WISE} observations; we have listed them in the table but have not extrapolated backward and forward, because there are not enough data to be confident in the comparison between $W1$, $W2$, and $V$-band. Tables \ref{tab:everything} and \ref{tab:xrays} list {\it all} of our observations over the entire course of the eruption, including those taken during or near eclipse, with Table \ref{tab:everything} giving all of the IR to UV magnitudes and Table \ref{tab:xrays} showing the X-ray count rates.

\section{Universal Decline Law}
\label{sec:udl}
\begin{figure*}
\centering
\plotone{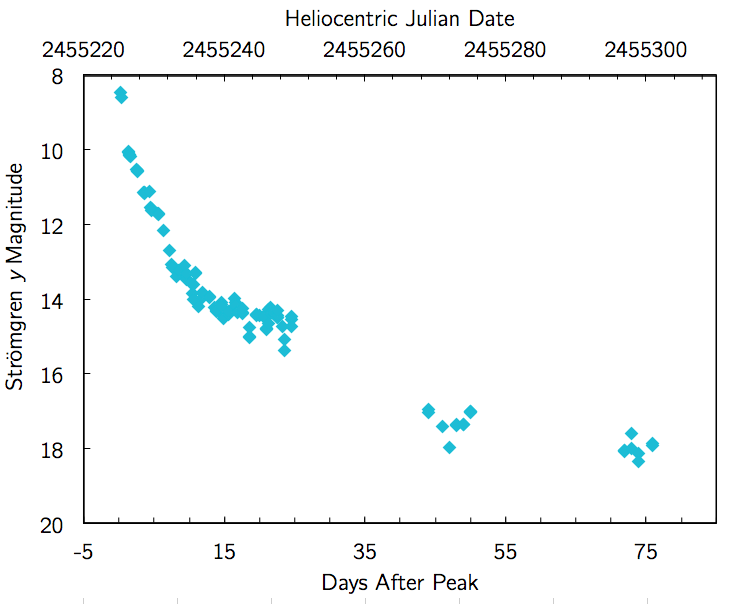}
\caption{Str\"omgren $y$ Light Curve, including all phases and therefore showing some eclipses. Str\"omgren $y$-band observations of U Sco during the 2010 eruption are from Handler on the SAAO 0.5m, Landolt on the KPNO 2.1m, Clem on the CTIO 1.0m, Kiyota using his personal 0.25m Schmidt-Cassegrain telescope, and Maehara using the Kwasan Observatory 0.25m. This filter is no longer available at many telescopes, so we were unable to get complete coverage throughout the eruption, but we were able to get some epochs. Str\"omgren $y$ is a narrow bandpass which is free from contamination by bright emission lines and is therefore ideal for testing \citet{hachisu2006a}'s universal decline law (Figure \ref{fig:yudl}).}
\label{fig:stromgreny}
\end{figure*}

\begin{figure*}
\centering
\epsscale{1.0}
\plotone{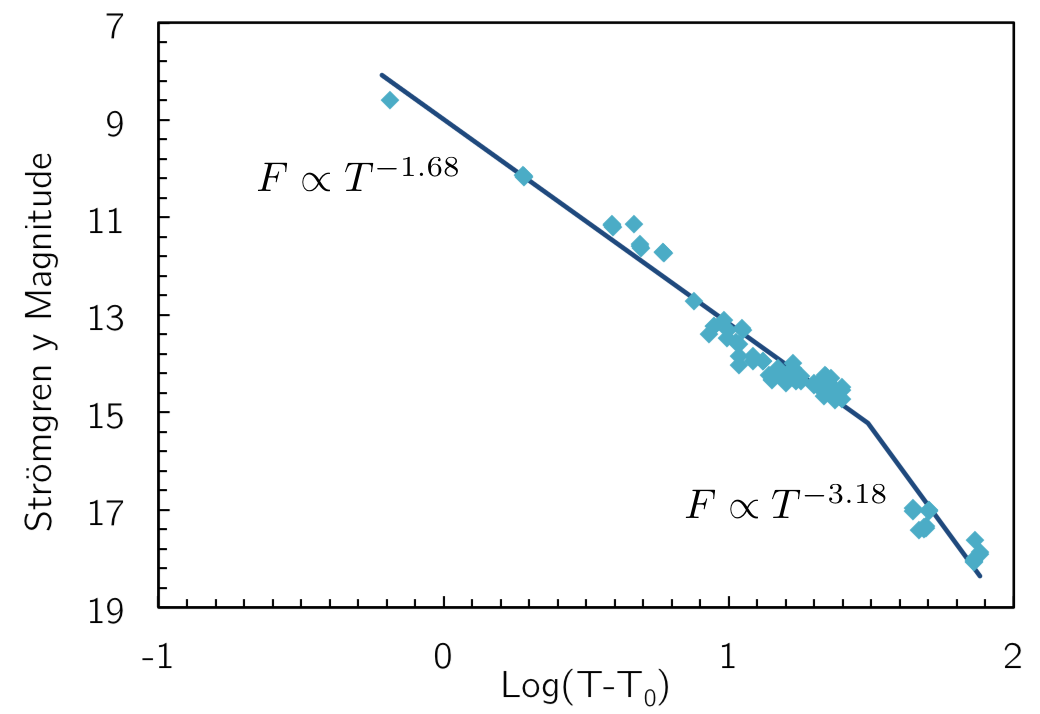}
\caption{Str\"omgren $y$ Universal Decline Law. Str\"omgren $y$ light curve (phases 0.1 to 0.9) of the U Sco 2010 eruption, plotted against the log of days after peak (Log($T - T_\mathrm{0}$)), and the best fit power laws to the data. The first power law has a best fit index of $-1.68 \pm 0.03$, and the second has a best fit power law index of $-3.18 \pm 0.10$. These are close to the predictions of \citet{hachisu2006a}, who give template fits of -1.75 and -3.5 for the first and second power laws, respectively.}
\label{fig:yudl}
\end{figure*}

\begin{figure*}
\centering
\epsscale{1.0}
\plotone{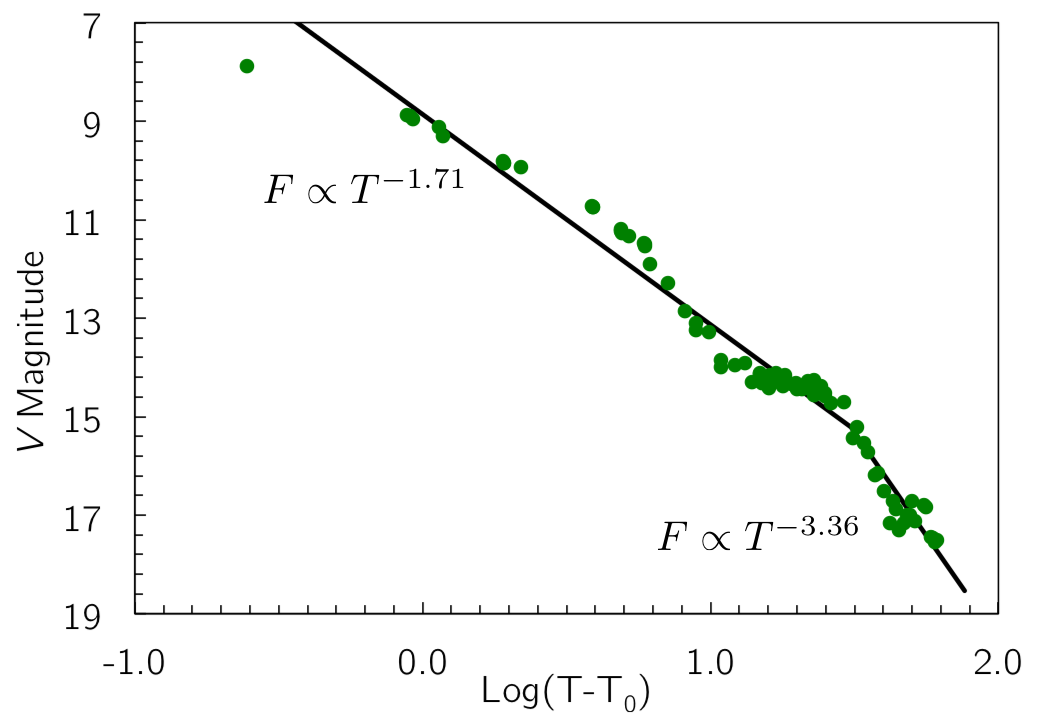}
\caption{$V$-Band Universal Decline Law. We also fit power laws to the $V$-band light curve, again plotted against the log of days after peak (Log($T - T_\mathrm{0}$)), which give best fit indices of $-1.71 \pm 0.02$ and $-3.36 \pm 0.14$ for the first and second parts of thelan light curve, respectively. This is nearly consistent, within 1$\sigma$ errors, with the predictions of \citet{hachisu2006a}. Although $V$-band does have potential emission line contamination, Figure \ref{fig:yvband} shows that it tracks well with Str\"omgren $y$ and therefore the power law fits are still useful. The advantage of using $V$-band is that we have more comprehensive coverage, which allows us to better constrain the break time.}
\label{fig:vudl}
\end{figure*}

\begin{figure*}
\centering
\epsscale{1.0}
\plotone{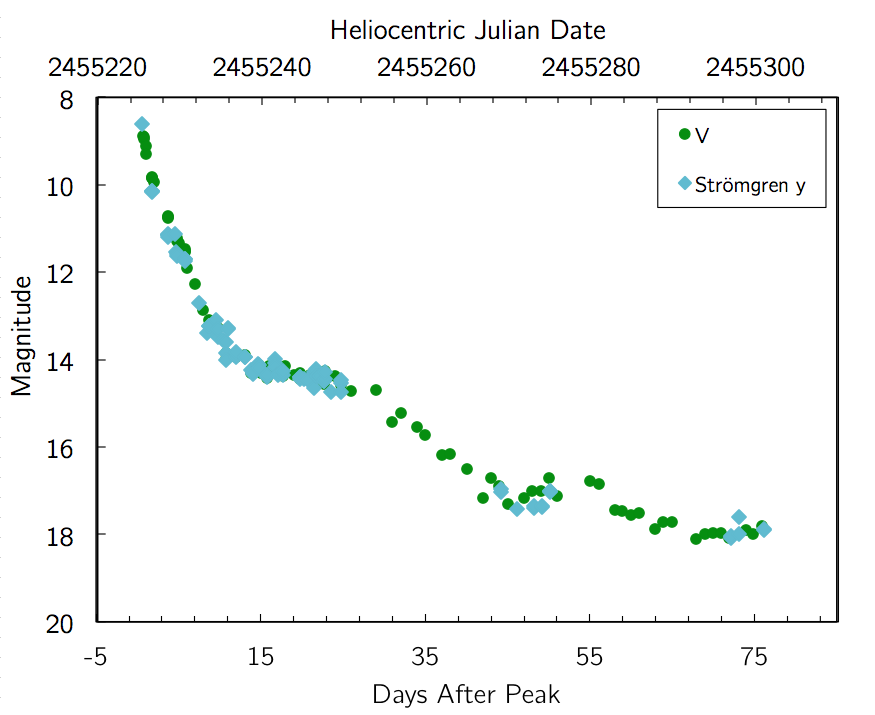}
\caption{Str\"omgren $y$ vs. $V$-Band. This comparison of the Str\"omgren $y$ and $V$-band light curves shows that they are nearly identical, despite the extra emission line flux in the $V$-band. For this reason, we are confident that it is reasonable to fit the $V$-band light curve with a broken power law (Figure \ref{fig:vudl}) to test the predictions of \citet{hachisu2006a}'s universal decline law.}
\label{fig:yvband}
\end{figure*}

\citet{hachisu2006a} introduced a universal decline law for novae that can be used to predict the turn-on and turn-off of supersoft X-ray flux as well as estimate the mass of the WD in some cases. Their template light curve has a slope of $F \sim t^{-1.75}$ shortly after peak, where $F$ is flux (measured in magnitudes) and $t$ is time (in days from peak), and a slope of $F \sim t^{-3.5}$ in the later part of the eruption light curve. They note that it is best to look at the narrow Str\"{o}mgren $y$-band light curve to test this law, because it is free of contamination from strong emission lines. We therefore arranged for as many Str\"{o}mgren $y$ observations of U Sco as possible. This is difficult because Str\"{o}mgren $y$ is no longer a common filter, but we were able to obtain coverage from Handler on the SAAO 0.5m, Landolt on the KPNO 2.1m, Clem on the CTIO 1.0m, Kiyota using his personal 0.25m Schmidt-Cassegrain telescope, and Maehara using the Kwasan Observatory 0.25m. The full Str\"{o}mgren $y$ light curve can be seen in Figure \ref{fig:stromgreny}. Figure \ref{fig:yudl} shows the best fit power laws for the Str\"{o}mgren $y$ data (using only the data from phases 0.1 to 0.9 to avoid the eclipses), with indices and $1\sigma$ errors of $-1.68 \pm 0.03$ for the first power law and $-3.18 \pm 0.10$ for the second. However, since the coverage is somewhat sparse, the break time is not well defined. To remedy this, we also fit the $V$-band data, shown in Figure \ref{fig:vudl}, which gives power law indices of $-1.71 \pm 0.02$ and $-3.36 \pm 0.14$ for the first and second power laws, respectively, and a break time of $\log(T-T_0)=1.487 \pm 0.005$, which gives $T=\textrm{HJD } 2455255.38 \pm 1.01$ days. Despite the fact that $V$ is a broad-band filter which includes some contaminating emission line fluxes, it is reasonable to fit the $V$-band data as well because $V$ and Str\"{o}mgren $y$ track each other almost identically, as shown in Figure \ref{fig:yvband}. We then used the break time found from the $V$ band data to better determine the slopes of the Str\"{o}mgren $y$ fits, and from that we obtained the fit values quoted above. For both bands, our results are nearly consistent with the predictions of \citet{hachisu2006a} within errors, with the small differences possibly due to the unusual features of the U Sco light curve such as the two plateaus.

For some novae, if they fit the model well, the break time can be used to determine the composition of the WD, whether it is carbon-oxygen or oxygen-neon-magnesium. Unfortunately for our case with U Sco, the models do not yet account for WDs very close to the Chandrasekhar limit, nor are they fully capable of dealing with RNe (I. Hachisu 2010, private communication), so we are unable to use this method to reliably determine the type of WD in the U Sco system. A crude extrapolation of \citet{hachisu2006a}'s Table 10 indicates that U Sco, with a $T-T_0 = 30.7$, should have a WD with a high neon content, but this is not a conclusive result.

\section{Spectral Energy Distribution}
\label{sec:sed}
\begin{figure*}
\centering
\epsscale{1.0}
\plotone{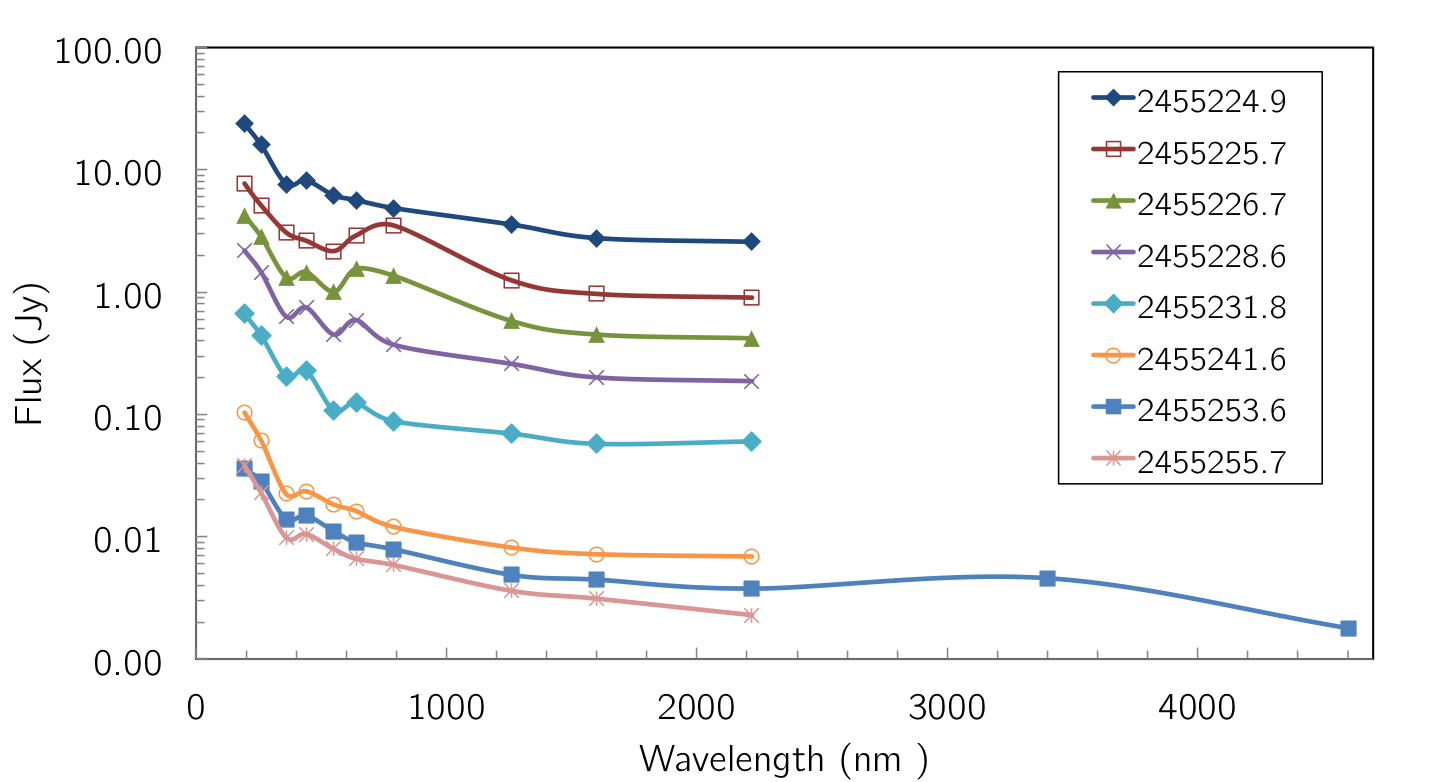}
\caption{Spectral Energy Distributions (SEDs) for selected days (HJD listed in the legend) during the 2010 eruption of U Sco. The wavelength range extends from the {\it Swift} UVOT $w2$ filter (192.8 nm) to $K$-band (2200 nm) for most of the SEDs; on HJD 2455253.6 (Day +29) we also had serendipitous {\it WISE} $W1$ and $W2$ observations, so those are included for that day. (The SED for 2455252.3, which also had {\it WISE} observations, is essentially identical to that of day 2455253.6, so we only include one.) The legend in the figure gives the mean HJD of the observations that went into constructing that particular distribution. The majority of the energy is released early in the eruption near the short end of the spectrum, so the UVOT $w1$ and $w2$ coverage is particularly valuable. There were essentially zero usable UVOT observations just after the peak on JD 2455224.7, so the shortest wavelength points on days 2455224.9-2455236.8 were extrapolated from the $B$ magnitudes on those dates as described in Section \ref{sec:sed}. The daily SEDs are used to calculate the total radiated energy, $E_\mathrm{rad}=6.99^{+0.83}_{-0.57} \times 10^{44}$ erg, which is then used to estimate the total amount of mass ejected during the eruption, $m_\mathrm{ej}=2.10^{+0.24}_{-0.17} \times 10^{-6} M_\odot$.}
\label{fig:seds}
\end{figure*}

With our comprehensive data set, we can construct daily SEDs for the entire eruption, allowing us to closely follow the multi-wavelength evolution of U Sco. The SEDs were constructed using predominantly the CTIO 1.3m, SAAO 0.5m, and {\it Swift} UVOT observations, so the average wavelength coverage is from 193 nm ({\it Swift} UVOT {\it w2}) to 2200 nm ({\it K}-band). Selected SEDs can be seen in Figure \ref{fig:seds}, which includes scientifically interesting days such as those just after peak (when most of the energy is released, in a wavelength range that unfortunately coincides with a gap in our coverage) and one of the days with {\it WISE} coverage, as well as more average days later in the eruption. Because of the scarcity of the {\it WISE} data and our lack of confidence in any extrapolation over the entire length of the eruption, the far-IR observations from {\it WISE} are only included on the actual days they were made. Since the amount of energy released in the far-IR is negligible compared to that released in shorter wavelengths, the inclusion of two days of {\it WISE} data does not introduce any measurable errors to the overall calculation. This is the first time any nova has had enough observations over a wide range of wavelengths and at such a fast time cadence that a plot such as this could be constructed.

Using a distance of 12 kpc \citep{schaefer2010b}, the average luminosity of the system in erg s$^{-1}$ can be calculated on a daily basis and then summed to obtain the total amount of radiated energy, $E_\mathrm{rad}$. We start with the observed magnitudes available for each HJD and then bin each day's observations to get a daily average. On days where there was no observation in a given band, the average luminosity was linearly interpolated based on the values for that band on the days immediately preceding and following it, or linearly extrapolated based on its average difference from a neighboring band for which there was an observation. The latter extrapolation is reasonable since the colors during eruption are relatively constant, as discussed in Section \ref{sec:observations}. For bands with shorter wavelengths ({\it Swift} {\it w1} and {\it w2} and Johnson {\it U}-band), we extrapolated based on the average {\it B}-band magnitude for that day, because {\it B}-band observations were made frequently and come closest to representing the amount of energy emitted in the near- and far-UV. For longer wavelength filters ({\it R$_c$I$_c$JHK}), we extrapolated from the average {\it V}-band magnitude for the given day. The interpolated and extrapolated magnitudes are denoted in Table \ref{tab:mags} with dagger ($\dagger$) and double dagger ($\ddagger$) symbols, respectively. We corrected most of the magnitudes for extinction using $A$ values from the NASA/IPAC IRSA Galactic Dust Reddening and Extinction service; the {\it Swift w1} and {\it w2} bands were corrected using the relation from \citet{cardelli1989a}, with $A_V = 0.74$, from an average $R_V=3.1$ and $E$($B-V$)=0.240 \citep{webbink1987a}. 

The majority of the energy of the eruption is released shortly after peak and is very blue, gaining in strength all the way through to the near UV, seen by {\it Swift}. Our coverage of this regime is not as thorough as we would like it to be in either wavelength range or time; in an ideal world we would have observations at wavelengths shorter than the two {\it Swift} bands as well as UV observations starting immediately after eruption. Although {\it Swift} did observe U Sco at very early times, UVOT photometry is unavailable from those times due to coincidence loss from the bright source. Although there is a possibility that a new analysis method may be able to be used to extract some information from these observations \citep{page2013b}, that is not currently an option. Instead we extrapolate to obtain sensible values for early UV magnitudes and acknowledge that we may be missing some of the energy being released in the nova, which introduces an unknown amount of systematic error. Overall, though, the extrapolation to near UV from the early {\it B}-band magnitudes is reasonable.

The daily magnitude averages were then converted to Janskys and the quiescent flux level in each band, based on pre-2010-eruption photometry, was subtracted out of each daily average to isolate the flux due to the eruption. Each day was then summed across all bandpasses to give the average total eruption luminosity, $L_\mathrm{tot}$, for each day. The $L_\mathrm{tot}$ values were multiplied by the duration (in seconds) of the time from whence those data came (where some of the ``days" are longer or shorter than 24 hr, since the timestamp given to each daily luminosity value is calculated as the average of the times of the observations that went into that day's average magnitudes) to give a daily total of radiated energy, and then finally each of those daily energies were summed to obtain the total amount of energy radiated during the eruption, $E_\mathrm{rad}$. For U Sco, $E_\mathrm{rad}=6.99 \times 10^{44}$ erg. Table \ref{tab:luminosities} gives the average daily luminosities for each band as well as the summed total for each day. 

We then use the method of \citet{shara2010a} to estimate the total amount of mass ejected during the eruption. This mass measurement is difficult to make; many previously published methods rely on a number of assumptions and are uncertain to orders of magnitude (see below for further discussion of these methods). The \citet{shara2010a} method was developed by searching their extensive grid of nova models for characteristics that correlate with ejected mass, $m_\mathrm{ej}$, under the important assumption that there is not equipartition between kinetic and radiated energy. They found a direct correlation between $m_\mathrm{ej}$ and $E_\mathrm{rad}$. Obtaining $E_\mathrm{rad}$ requires comprehensive coverage of the eruption in multiple wavelengths and, for the first time, we have exactly what is needed. Although theory predicts that most RNe should see a net WD mass gain over their eruption cycles \citep{yaron2005a}, we seek observational proof, and the \citet{shara2010a} method provides a new and independent way to evaluate this.

Following \citet{shara2010a}, we can use $E_\mathrm{rad}$ to obtain the total $m_\mathrm{ej}$ as
\begin{equation}
m_\mathrm{ej}=6 \times 10^{-18} E_\mathrm{rad}
\end{equation}
where $m_\mathrm{ej}$ is measured in g and $E_\mathrm{rad}$ in erg. With this, our estimate of the total mass ejected during the 2010 eruption is $m_\mathrm{ej}=4.20 \times 10^{27}$ g $= 2.10 \times 10^{-6} M_\odot$. 

It is difficult to place error bars on these values, as the uncertainties are not well defined, but we can make a good estimate. The majority of the radiated energy comes out right at the peak of the eruption in the UV regime. As noted previously, the early {\it Swift} observations are not usable, so the UVOT $w1$ and $w2$ values just after peak are extrapolated from the $B$-band measurement and model predictions. To estimate the error on the $m_\mathrm{ej}$ measurement, we vary the first extrapolated $w2$ value (which represents the majority of the energy released in the eruption) by one standard deviation of the ${<}B-w2{>}$ values for times at which both were measured. This gives $1\sigma$ error bars on the $E_\mathrm{rad}$ of $+0.83 \times 10^{44}$ erg and $-0.57 \times 10^{44}$ erg, for a final $E_\mathrm{rad} = 6.99^{+0.83}_{-0.57} \times 10^{44}$ erg. For the ejected mass, the $1\sigma$ errors are $+0.24 \times 10^{-6} M_\odot$ and $-0.17 \times 10^{-6} M_\odot$. We thus conclude that $m_\mathrm{ej}=2.10^{+0.24}_{-0.17} \times 10^{-6} M_\odot$. 

Our new value of $m_\mathrm{ej}=2.10^{+0.24}_{-0.17} \times 10^{-6} M_\odot$ is consistent with one of the (highly uncertain) measurements from previous eruptions, ${\sim} 3 \times 10^{-6} M_\odot$ from \citet{hachisu2000a}, but an order of magnitude larger than the ${\sim}  10^{-7} M_\odot$ calculated by \citet{anupama2000a}. \citet{hachisu2000a} obtain their value of $m_\mathrm{ej}$ from light curve modeling, while \citet{anupama2000a} consider the flux in the Balmer lines 11-12 days after peak. Both methods rely on a number of assumptions, detailed extensively in Appendix A of \citet{schaefer2011a}, which add significant uncertainties and model-dependencies to the results. Although the uncertainties on our $m_\mathrm{ej}$ value are not as well understood as we would like, the lack of dependence on any particular model gives us more confidence in our value than in these previous estimates. 

Because U Sco is an eclipsing binary, the orbital period change across each eruption ({$\Delta P$}) can be precisely calculated by measuring the change in eclipse times. The orbital period change across a given eruption and the amount of mass ejected during that eruption are related by $m_{\mathrm{ej}}=(M_{\mathrm{WD}}/A)({\Delta} P/P)$, where {$A$} is a parameter that depends on the fraction of ejected matter captured by the companion and the specific angular momentum of the ejecta  \citep{schaefer1983a}. \cite{schaefer2011a} carefully measured {$\Delta P$} across the 1999 eruption and, using the aforementioned relationship, derived $m_\mathrm{ej}=(4.3 \pm 6.7) \times 10^{-6} M_\odot$, which is consistent with our result. Using the same method for the 2010 eruption, however, results from \cite{schaefer2013a} indicate that U Sco ejected $2.5 \times 10^{-5} M_{\odot}$ in its most recent eruption, which does not agree with our conclusions in this manuscript by an order of magnitude.

\section{Net Mass Change of the U Sco White Dwarf}
\label{sec:mass}
The $m_\mathrm{ej}$ value is crucial for determining whether U Sco and similar RNe can become SNe Ia. For the WD to reach the Chandrasekhar limit and explode, it must have a net mass gain over its lifetime. We can compare the $m_\mathrm{ej}$ measurement with the total amount of mass accreted during the time preceding the eruption, $m_\mathrm{acc}$, to determine the overall mass change of the WD. Although in principle it is straightforward to measure the accreted mass by considering the $B$-band flux, in reality there are a number of uncertainties. The following values have been proposed for the yearly accretion rate of U Sco, all in units of $M_\odot$ yr$^{-1}$: $<3 \times 10^{-7}$ \citep{shen2007a}, $4 \times 10^{-7}$ \citep{duschl1990a}, $2.5 \times 10^{-7}$ \citep{hachisu2000a}, $4.4 \times 10^{-7}$ \citep{starrfield1988a}, and $1 \times 10^{-7}$ \citep{kato1990a}. $\dot{M}$ is therefore likely in the range of $1 - 4.4 \times 10^{-7} M_\odot$ yr$^{-1}$. For an average recurrence time of 10 years, this means that the total $m_\mathrm{acc}$ is therefore in the range $1-4.4 \times 10^{-6} M_\odot$ with an average value and 1$\sigma$ error of $(2.9 \pm 1.3) \times 10^{-6}$ $M_\odot$. Unfortunately, the values of $m_\mathrm{ej}$ and $m_\mathrm{acc}$ are too close, and the error bars too large, to make a conclusive statement about the net mass change of U Sco. An increase, decrease, or no change at all in the mass of the U Sco WD would all be consistent with this result.

\section{Conclusions}
The 2010 eruption of U Sco was the all-time best observed nova eruption at the time. (In sheer number of observations, U Sco has since been eclipsed by the 2011 eruption of the RN T Pyxidis \citep{schaefer2012b}, but U Sco still has the most comprehensive overall coverage.) The early discovery, fast notification, and regular observations by our collaboration of professional and  amateur astronomers allowed us to define the entire eruption light curve from shortly after peak until the return to quiescence 67 days later. For previous eruptions there are a few isolated points late in the tail of the decline, but this was the first time U Sco was intentionally monitored for more than 30 days after peak \citep{schaefer2010b}. This systematic coverage allows us to describe the overall shape of the light curve, starting with the fast decline (days 0-14), and continuing through the first plateau (days 14-32) which is caused by the supersoft X-ray source as described by \citet{hachisu2008a} and coincides with the return of the optical eclipses. The first plateau falls off starting on day 32, and our long-term coverage of the eruption allowed us to discover the second plateau in the light curve, which lasts from days 41 to 54 and is as yet unexplained, but occurs at the same time as the aperiodic optical dips that have been linked to transient vertical structure in the accretion disk as it reforms after being blown away during the eruption \citep{pagnotta2010a,schaefer2011b}. Following the second plateau, U Sco undergoes a jittery return to quiescence (days 54-67), where it will remain until the next eruption, which should occur around 2020.

We were able to combine our comprehensive $V$-band coverage with our Str\"{o}mgren $y$ observations to show that U Sco is nearly consistent (within $1\sigma$ errors) with the universal decline law of  \citet{hachisu2006a}. We were unable to obtain a high-enough cadence in $y$-band to use it exclusively as recommended, but our multiple Str\"{o}mgren $y$ observations at different epochs allow us to compare it with the $V$-band light curve and see that they track each other nearly identically, giving us confidence that we can use the $V$-band data to test the universal decline law. We calculate power law indices of $-1.71 \pm 0.02$ for the first phase and $-3.36 \pm 0.14$ for the second, and a break time of $T=\textrm{HJD } 2455255.38 \pm 1.01$ days. Our fits our nearly consistent with the template values from \citet{hachisu2006a} of $-1.75$ and $-3.5$ for the first and second power laws, respectively, with the first power law falling just outside the $1\sigma$ range of our fit, and the second falling just on the edge. This is likely due to the fact that U Sco does not have a smooth light curve shape, but instead has the unusual features detailed above, which slightly skew the light curve away from the model, since it is constructed for novae with smoother, more regular declines.

Our full multi-wavelength coverage confirmed the predicted temporal correlation between the (first) optical/near-IR plateau, the return of the eclipses, and the detection of the supersoft X-rays. The simultaneity of these three events is due to the shell becoming optically thin at that time, revealing the inner binary, which allows us to see both the X-rays and the reestablishment of the eclipses. The high cadence UV coverage shows for the first time how well the UV light curve tracks the optical one, including showing the primary eclipses and possibly the secondary eclipses as well.

We were also able, for the first time, to construct daily SEDs of a nova eruption covering an average wavelength range of 193 nm ({\it Swift} UVOT {\it w2}) to 2200 nm ({\it K}-band). Using this unique data set, we were able to calculate the total radiated energy from the eruption, $E_\mathrm{rad}=6.99^{+0.83}_{-0.57} \times 10^{44}$ erg, which we then used to estimate the amount of mass ejected, $m_\mathrm{ej}=2.05^{+0.26}_{-0.18} \times 10^{-6} M_\odot$, following \cite{shara2010a}. This calculation would not have been possible without the exceptional amount of data that was collected for U Sco throughout the entire eruption over many wavelengths. Unfortunately, the measurements of the total mass accreted since the last eruption, $m_\mathrm{acc}$, are not precise enough to allow us to determine whether the U Sco WD is gaining or losing mass over the course of a full eruption cycle. In spite of this, the \citet{shara2010a} method is an improvement upon previous methods and should be tested on and applied to future nova eruptions whenever enough observations can be made. Early UV coverage is particularly important to this computation, and this can best be obtained from the {\it Swift} UVOT as soon as possible after the eruptions are detected.

This is the most comprehensive multi-wavelength light curve of a nova eruption, covering more wavelengths over a longer time period than ever before. With this unprecedented data set we discovered three new phenomena and tested some of the theoretical predictions that have been made for novae in general and U Sco in particular. We encourage future similar campaigns whenever possible to determine whether these new phenomena are unique to U Sco or are in fact common among novae or recurrent novae, and to continue to find unexpected phenomena as we open new time and wavelength regimes.

\acknowledgments
This research was supported by NSF Grant AST-0708079, the Louisiana Space Consortium and NASA through Grant NNX10AI40H, as well as the Kathryn W. Davis Postdoctoral Scholar program, which is supported in part by the New York State Education Department and by the National Science Foundation under grant numbers DRL-1119444 and DUE-1340006. AUL acknowledges support from NSF Grant AST-0803158. KLP and JPO acknowledge support from the UK Space Agency. GH acknowledges partial financial support from the Austrian FWF grant P20526-N16 and encouragement from Patrick Woudt.

We heartily thank all members of the USCO2010 collaboration, whose observations were critical to our successful campaign. We acknowledge with many thanks the variable star observations from the AAVSO International Database contributed by observers worldwide and used in this research, as well as the observers of the Center for Backyard Astrophysics. This publication makes use of data products from the {\it Wide-field Infrared Survey Explorer}, which is a joint project of the University of California, Los Angeles, and the Jet Propulsion Laboratory/California Institute of Technology, funded by the National Aeronautics and Space Administration. This research has made use of the NASA/ IPAC Infrared Science Archive, which is operated by the Jet Propulsion Laboratory, California Institute of Technology, under contract with the National Aeronautics and Space Administration.

\clearpage
\LongTables

\begin{centering}
\begin{deluxetable}{lllll}
\tabletypesize{\tiny}
\tablecaption{Observer Details}
\tablewidth{0pt}
\setlength{\tabcolsep}{0.05in} 
\tablehead{
\colhead{Observer}   &
\colhead{Telescope}   &
\colhead{Location}   &
\colhead{Filters}   &
}
\startdata
Pagnotta	&	CTIO 1.3m	&	Cerro Tololo, Chile	&	{\it BVRIJHK}			\\
Handler	&	SAAO 0.5m	&	Sutherland, South Africa	&	{\it UBVRIy}			\\
Clem	&	CTIO 1.3m	&	Cerro Tololo, Chile	&	Str\"{o}mgren $y$			\\
Landolt	&	KPNO 2.1m	&	Kitt Peak, Arizona	&	Str\"{o}mgren $y$			\\
Maehara	&	Kwasan 0.25m	&	Kyoto, Japan	&	Str\"{o}mgren $y$		\\
Kiyota	&	0.25m Schmidt-Cassegrain	&	Ibaraki, Japan	&	Str\"{o}mgren $y$			\\
...	&	{\it Swift} UVOT	&	...	&	$w1$ (260 nm), $w2$ (193 nm)			\\
...	&	{\it Swift} XRT	&	...	&	0.3-10 keV			\\
...	&	{\it WISE}	&	...	&	$W1$ (3400 nm), $W2$ (4600 nm)			\\
\enddata
\label{tab:observers}
\end{deluxetable}
\end{centering}

\begin{centering}
\begin{deluxetable}{lllllllllllcc}
\tabletypesize{\tiny}
\tablecaption{Average Daily Magnitudes\tablenotemark{a}}
\tablewidth{0pt}
\setlength{\tabcolsep}{0.05in} 
\tablehead{
\colhead{HJD}   &
\colhead{{\it Swift} $w2$}   &
\colhead{{\it Swift} $w1$}   &
\colhead{$U$}   &
\colhead{$B$}   &
\colhead{$V$}   &
\colhead{$R_c$}   &
\colhead{$I_c$}   &
\colhead{$J$}   &
\colhead{$H$}   &
\colhead{$K$}   &
\colhead{{\it WISE} $W1$} 	&
\colhead{{\it WISE} $W2$} 
}
\startdata
2455224.90	&	$	7.40	^\ddagger$	&	$	7.38	^\ddagger$	&	$	7.42	^\ddagger$	&	$	8.10	^\ddagger$	&		7.89		&	$	7.60	^\ddagger$	&	$	7.34	^\ddagger$	&	$	6.88	^\ddagger$	&	$	6.64	^\ddagger$	&	$	6.15	^\ddagger$	&	\ldots	&	\ldots	\\
2455225.69	&	$	8.64	^\ddagger$	&	$	8.62	^\ddagger$	&		8.41		&		9.34		&		9.03		&		8.32		&		7.70		&	$	8.02	^\ddagger$	&	$	7.78	^\ddagger$	&	$	7.29	^\ddagger$	&	\ldots	&	\ldots	\\
2455226.66	&	$	9.28	^\ddagger$	&	$	9.26	^\ddagger$	&		9.32		&		9.98		&		9.86		&		9.01		&		8.73		&	$	8.86	^\ddagger$	&	$	8.61	^\ddagger$	&	$	8.12	^\ddagger$	&	\ldots	&	\ldots	\\
2455227.63	&	$	9.64	^\ddagger$	&	$	9.62	^\ddagger$	&	$	9.72	^\dagger$	&	$	10.34	^\dagger$	&	$	10.30	^\dagger$	&	$	9.53	^\dagger$	&	$	9.43	^\dagger$	&	$	9.29	^\ddagger$	&	$	9.05	^\ddagger$	&	$	8.56	^\ddagger$	&	\ldots	&	\ldots	\\
2455228.59	&	$	10.00	^\ddagger$	&	$	9.98	^\ddagger$	&		10.13		&		10.70		&		10.74		&		10.06		&		10.14		&	$	9.73	^\ddagger$	&	$	9.49	^\ddagger$	&	$	8.99	^\ddagger$	&	\ldots	&	\ldots	\\
2455229.69	&	$	10.40	^\ddagger$	&	$	10.38	^\ddagger$	&		10.60		&		11.09		&		11.26		&		10.62		&		10.67		&		10.12		&		9.83		&		9.37		&	\ldots	&	\ldots	\\
2455230.68	&	$	10.72	^\ddagger$	&	$	10.70	^\ddagger$	&		10.82		&		11.42		&		11.61		&		11.07		&		11.06		&		10.60		&		10.30		&		9.66		&	\ldots	&	\ldots	\\
2455231.83	&	$	11.30	^\ddagger$	&	$	11.27	^\ddagger$	&	$	11.35	^\dagger$	&		11.99		&		12.28		&		11.75		&		11.70		&		11.15		&		10.84		&		10.22		&	\ldots	&	\ldots	\\
\enddata
\label{tab:mags}
\tablenotetext{a}{Interpolated values are denoted with a single dagger ($\dagger$); extrapolated values are denoted with a double dagger ($\ddagger$). See Section \ref{sec:sed} for more details.}
\end{deluxetable}
\end{centering}

\begin{centering}
\begin{deluxetable}{llccccr}
\tabletypesize{\tiny}
\tablecaption{U Sco 2010 Observations, Infrared to Ultraviolet Magnitudes}
\tablewidth{0pt}
\setlength{\tabcolsep}{0.05in} 
\tablehead{
\colhead{HJD}   &
\colhead{Phase}   &
\colhead{Observer}   &
\colhead{Telescope}   &
\colhead{Filter}   &
\colhead{Magnitude}   &
\colhead{Error}   
}
\startdata
2455224.9352	&	0.795	&	Harris	&	0.4m Schmidt-Cassegrain	&	{\it V}	&	7.89	&	0.01	\\
2455225.2886	&	0.082	&	Maehara	&	Kwasan 0.25m	&	Stromgren {\it y}	&	8.47	&	0.02	\\
2455225.3382	&	0.122	&	Kiyota	&	0.25m Schmidt-Cassegrain	&	Stromgren {\it y}	&	8.60	&	0.02	\\
2455225.5749	&	0.314	&	Handler	&	SAAO 0.5m	&	{\it V}	&	8.88	&	0.02	\\
2455225.5749	&	0.314	&	Handler	&	SAAO 0.5m	&	$I_c$	&	7.58	&	0.02	\\
2455225.5749	&	0.314	&	Handler	&	SAAO 0.5m	&	{\it B}	&	9.23	&	0.02	\\
2455225.5907	&	0.327	&	Handler	&	SAAO 0.5m	&	{\it V}	&	8.90	&	0.02	\\
2455225.5907	&	0.327	&	Handler	&	SAAO 0.5m	&	{\it U}	&	8.38	&	0.02	\\
\enddata
\label{tab:everything}
\end{deluxetable}
\end{centering}

\begin{centering}
\begin{deluxetable}{llcl}
\tabletypesize{\tiny}
\tablecaption{U Sco 2010 Observations, X-Ray Count Rates}
\tablewidth{0pt}
\setlength{\tabcolsep}{0.05in} 
\tablehead{
\colhead{HJD}   &
\colhead{Phase}   &
\colhead{Count Rate}   &
\colhead{Error}   
}
\startdata
2455232.5034	&	0.9449	&	0.0015	&	0.0005	\\
2455229.9981	&	0.9090	&	0.0019	&	0.0004	\\
2455228.4198	&	0.6264	&	0.0025	&	0.0003	\\
2455282.7612	&	0.7867	&	0.0045	&	0.0012	\\
2455231.1363	&	0.8340	&	0.0046	&	0.0010	\\
2455287.9803	&	0.0280	&	0.0068	&	0.0016	\\
2455287.2818	&	0.4604	&	0.0071	&	0.0014	\\
2455284.6330	&	0.3079	&	0.0072	&	0.0015	\\
\enddata
\label{tab:xrays}
\end{deluxetable}
\end{centering}

\clearpage
\LongTables
\begin{landscape}
\begin{centering}
\begin{deluxetable}{lllllllllllccl}
\tabletypesize{\tiny}
\tablecaption{Average Daily U Sco Luminosities in erg s$^{-1}$}
\tablewidth{0pt}
\setlength{\tabcolsep}{0.04in} 
\tablehead{
\colhead{HJD}   &
\colhead{{\it {\it Swift}} $w2$}   &
\colhead{{\it Swift} $w1$}   &
\colhead{$U$}   &
\colhead{$B$}   &
\colhead{$V$}   &
\colhead{$R_c$}   &
\colhead{$I_c$}   &
\colhead{$J$}   &
\colhead{$H$}   &
\colhead{$K$}   &
\colhead{{\it WISE} $W1$}   &
\colhead{{\it WISE} $W2$}   &
\colhead{$L_\mathrm{total}$} 
}
\startdata
2455224.90	&	2.60E+39	&	1.02E+39	&	3.10E+38	&	2.03E+38	&	1.13E+38	&	8.06E+37	&	9.65E+37	&	5.90E+37	&	2.45E+37	&	2.20E+37	&	\ldots	&	\ldots	&	4.53E+39	\\
2455225.69	&	8.34E+38	&	3.28E+38	&	1.25E+38	&	6.52E+37	&	3.97E+37	&	4.15E+37	&	6.95E+37	&	2.06E+37	&	8.56E+36	&	7.69E+36	&	\ldots	&	\ldots	&	1.54E+39	\\
2455226.66	&	4.62E+38	&	1.82E+38	&	5.36E+37	&	3.61E+37	&	1.85E+37	&	2.21E+37	&	2.70E+37	&	9.60E+36	&	3.98E+36	&	3.58E+36	&	\ldots	&	\ldots	&	8.19E+38	\\
2455227.63	&	3.31E+38	&	1.30E+38	&	3.70E+37	&	2.59E+37	&	1.23E+37	&	1.36E+37	&	1.41E+37	&	6.41E+36	&	2.66E+36	&	2.39E+36	&	\ldots	&	\ldots	&	5.76E+38	\\
2455228.59	&	2.37E+38	&	9.34E+37	&	2.56E+37	&	1.85E+37	&	8.24E+36	&	8.42E+36	&	7.36E+36	&	4.28E+36	&	1.78E+36	&	1.60E+36	&	\ldots	&	\ldots	&	4.06E+38	\\
2455229.69	&	1.65E+38	&	6.50E+37	&	1.65E+37	&	1.29E+37	&	5.09E+36	&	5.00E+36	&	4.50E+36	&	3.00E+36	&	1.30E+36	&	1.12E+36	&	\ldots	&	\ldots	&	2.79E+38	\\
2455230.68	&	1.23E+38	&	4.82E+37	&	1.34E+37	&	9.56E+36	&	3.69E+36	&	3.29E+36	&	3.15E+36	&	1.92E+36	&	8.36E+35	&	8.61E+35	&	\ldots	&	\ldots	&	2.08E+38	\\
2455231.83	&	7.22E+37	&	2.84E+37	&	8.29E+36	&	5.63E+36	&	1.98E+36	&	1.77E+36	&	1.75E+36	&	1.15E+36	&	5.09E+35	&	5.13E+35	&	\ldots	&	\ldots	&	1.22E+38	\\
\enddata
\label{tab:luminosities}
\tablenotetext{a}{0.00E+00 entries denote values so small they are essentially equal to 0.}
\end{deluxetable}
\end{centering}
\clearpage
\end{landscape}

\end{document}